\documentstyle[sprocl,epsf,psfig,colordvi]{article}

\newcommand{\nub}{\overline{\nu}}

\def\be{\begin{equation}}
\def\ee{\end{equation}}
\def\bea{\begin{eqnarray}}
\def\eea{\end{eqnarray}}

\def\nim#1#2#3  {{\em Nucl. Instr. Meth.} {\bf#1}, #2 (#3). }
\def\np#1#2#3   {{ Nucl. Phys.} {\bf#1}, #2 (#3). }
\def\pcps#1#2#3 {{ Proc. Cam. Phil. Soc.} {\bf#1}, #2 (#3). }
\def\pl#1#2#3   {{ Phys. Lett.} {\bf#1}, #2 (#3). }
\def\plc#1#2#3   {{ Phys. Lett.} {\bf#1}, #2 (#3); }
\def\prep#1#2#3 {{ Phys. Rep.} {\bf#1}, #2 (#3). }
\def\prev#1#2#3 {{ Phys. Rev.} {\bf#1}, #2 (#3). }
\def\prl#1#2#3  {{ Phys. Rev. Lett.} {\bf#1}, #2 (#3). }
\def\prs#1#2#3  {{ Proc. Roy. Soc.} {\bf#1}, #2 (#3). }
\def\ptp#1#2#3  {{ Prog. Th. Phys.} {\bf#1}, #2 (#3). }
\def\rmp#1#2#3  {{ Rev. Mod. Phys.} {\bf#1}, #2 (#3). }
\def\rpp#1#2#3  {{ Rep. Prog. Phys.} {\bf#1}, #2 (#3). }
\def\zp#1#2#3   {{ Z. Phys.} {\bf#1}, #2 (#3). }
\def\epj#1#2#3   {{ Eur. Phys. Jour.} {\bf#1}, #2 (#3). }

\begin{document}

\title{ EXTRACTION OF  $R=\frac{\sigma_L}{\sigma _T} $ FROM CCFR
  $\nu_\mu$-Fe and $\nub_\mu$-Fe DIFFERENTIAL CROSS SECTIONS}

\author{A.~Bodek,$^{7}$ T.~Adams,$^{4}$ A.~Alton,$^{4}$
C.~G.~Arroyo,$^{2}$ S.~Avvakumov,$^{7}$ L.~de~Barbaro,$^{5}$
P.~de~Barbaro,$^{7}$ A.~O.~Bazarko,$^{2}$ R.~H.~Bernstein,$^{3}$
 T.~Bolton,$^{4}$ J.~Brau,$^{6}$ D.~Buchholz,$^{5}$
H.~Budd,$^{7}$ L.~Bugel,$^{3}$ J.~Conrad,$^{2}$ R.~B.~Drucker,$^{6}$
B.~T.~Fleming,$^{2}$
J.~A.~Formaggio,$^{2}$ R.~Frey,$^{6}$ J.~Goldman,$^{4}$
M.~Goncharov,$^{4}$ D.~A.~Harris,$ ^{7} $ R.~A.~Johnson,$^{1}$
J.~H.~Kim,$^{2}$ B.~J.~King,$^{2}$ T.~Kinnel,$^{8}$
S.~Koutsoliotas,$^{2}$ M.~J.~Lamm,$^{3}$ W.~Marsh,$^{3}$
D.~Mason,$^{6}$ K.~S.~McFarland, $^{7}$ C.~McNulty,$^{2}$
S.~R.~Mishra,$^{2}$ D.~Naples,$^{4}$  P.~Nienaber,$^{3}$
A.~Romosan,$^{2}$ W.~K.~Sakumoto,$^{7}$ H.~Schellman,$^{5}$
F.~J.~Sciulli,$^{2}$ W.~G.~Seligman,$^{2}$ M.~H.~Shaevitz,$^{2}$
W.~H.~Smith,$^{8}$ P.~Spentzouris, $^{2}$ E.~G.~Stern,$^{2}$
N.~Suwonjandee,$^{1}$ A.~Vaitaitis,$^{2}$ M.~Vakili,$^{1}$U.~K.~Yang,$^{7}$   
J.~Yu,$^{3}$ G.~P.~Zeller,$^{5}$ and E.~D.~Zimmerman$^{2}$}

\address{(Presented by A. Bodek for the CCFR/NuTeV Collaboration ) \\
$^{1}$ University of Cincinnati, Cincinnati, OH 45221 \\
$^{2}$ Columbia University, New York, NY 10027 \\
$^{3}$ Fermi National Accelerator Laboratory, Batavia, IL 60510 \\
$^{4}$ Kansas State University, Manhattan, KS 66506 \\
$^{5}$ Northwestern University, Evanston, IL 60208 \\
$^{6}$ University of Oregon, Eugene, OR 97403 \\
$^{7}$ University of Rochester, Rochester, NY 14627 \\
$^{8}$ University of Wisconsin, Madison, WI 53706\\ }

\maketitle\abstracts{
We report on the extraction of $R=\frac{\sigma_L}{\sigma _T} $
from CCFR  $\nu_\mu$-Fe and $\nub_\mu$-Fe
differential cross sections.
$R$ as measured in $\nu_\mu$ scattering is in agreement
with $R$ as measured in muon  and electron scattering.
All data on $R$ for $Q^2 > 1$ GeV$^2$
are in agreement with a NNLO QCD calculation which uses NNLO PDFs and includes
target mass effects.
We report on the first measurements of  $R$
in the low $x$ and  $Q^2 < 1$ GeV$^2$ region (where an anomalous large
rise in $R$ for nuclear targets has been observed by the
HERMES collaboration).-- UR-1635, Proceedings of DIS2001, Bologna April 2001}
%



   The ratio of longitudinal and transverse structure function, 
  $R$ (=$F_L/2xF_1$) in deep
  inelastic lepton-nucleon scattering experiments is a sensitive
test of the quark parton model of the nucleon.  Recently, there has been a renewed
interest in $R$ at small values of $x$ and $Q^2$, because of the large
anomalous nuclear effect that has been reported by the
HERMES experiment~\cite{HERMES}. A large value of $R$ in nuclear
targets could be interpreted as evidence for non spin 1/2
constituents, such as $\rho$ mesons in nuclei~\cite{MIT}.
Previous measurements of $R$ in muon and electron scattering ($R^{\mu/e}$)
are well described by the $R_{world}^{\mu/e}$~\cite{RWORLD} QCD inspired
empirical fit.
The $R_{world}^{\mu/e}$ fit is also in good agreement
with recent NMC muon data
 for $R$ at low $x$, and
with theoretical predictions~\cite{dupaper}
  $R_{NNLO+TM}^{\mu/e}$
  (a Next to Next to Leading (NNLO) QCD calculation using NLO
  Parton Distribution Functions (PDFs), and including target mass effects).
Very recently the NNLO calculations for $F_{L}$ and $F_{2}$ have been
updated~\cite{MRSNNLO}
to include estimates of the contribution from NNLO PDFs. The quantity 
$R_{NNLOpdfs+TM}^{\mu/e}$ is extracted by adding target mass effects 
to these calculations of $F_{L}$. 
For $x>0.1$ it is expected that $R^\nu$  should
be the same as $R^{\mu/e}$. 
However, for $x<0.1$ and low $Q^2$  (in leading order),
$R^\nu$  is expected to be larger
than $R^{\mu/e}$ because of the production of
  massive charm quarks in the final state.
We calculate a correction to $R_{world}^{\mu/e}$ for this
difference  using a leading order slow rescaling model with a charm mass, $m_c (=1.3$ GeV)
and obtain an  effective
$R_{world}$ for $\nu_\mu$ scattering ($R_{eff}^{\nu}$). 
Here, we report on an extraction of $R$
in neutrino scattering ($R^\nu$), extending to low $x$ and $Q^2$,
ad compare
to $R^{\mu/e}$ data and to predictions from $R_{eff}^{\nu}$,
$R_{world}^{\mu/e}$, and $R_{NNLOpdfs+TM}^{\mu/e}$).

The sum of $\nu_\mu$ and $\nub_\mu$
  differential cross sections
for charged current interactions on isoscalar target is related to the
structure functions as follows:
\begin{tabbing}
$F(\epsilon)$ \= $\equiv \left[\frac{d^2\sigma^{\nu }}{dxdy}+
\frac{d^2\sigma^{\overline \nu}}{dxdy} \right]
  \frac {(1-\epsilon)\pi}{y^2G_F^2ME}$ \\
   \> $ = 2xF_1 [ 1+\epsilon R ] + \frac {y(1-y/2)}{1+(1-y)^2} \Delta 
xF_3 $. \hspace{0.7in} (1)
\end{tabbing}
Here $G_{F}$ is the weak Fermi coupling constant, $M$ is the nucleon
mass, $E_{\nu}$ is the incident energy, the scaling
variable $y=E_h/E_\nu$ is the fractional energy transferred to
the hadronic vertex, $E_h$ is the final state hadronic
energy, and $\epsilon\simeq2(1-y)/(1+(1-y)^2)$ is the polarization of virtual
$W$ boson. The structure
function $2xF_1$ is expressed in terms of $F_2$
by $2xF_1(x,Q^2)=F_2(x,Q^2)\times
\frac{1+4M^2x^2/Q^2}{1+R(x,Q^2)}$, where $Q^2$ is the
square of the four-momentum transfer to the nucleon,
   $x=Q^2/2ME_h$ (the Bjorken scaling variable) is
the fractional momentum carried by the struck quark,
and $R=\frac{\sigma _L}{\sigma _T} = \frac{F_L}{2xF_1}$.

Values of $R$ (or equivalently $F_L$)
and $2xF_1$ are extracted from the sums of
the corrected $\nu_\mu$-Fe and $\nub_\mu$-Fe
differential cross sections at different
energy bins according to Eq. (1).
An extraction of $R$ using Eq. (1) requires a knowledge of
$\Delta xF_3$ term, which in leading order
$\simeq4x(s-c)$. We obtain
$\Delta xF_3$ from  theoretical predictions~\cite{DXF3} for massive
charm production using the TR-VFS NLO calculation
with the extended MRST99 and the suggested scale $\mu=Q$.
This prediction is used as input to Eq. (1) in the extraction of $R^\nu$.
This model yields $\Delta xF_3$ values similar
to the NLO ACOT Variable Flavor Scheme
(implemented with CTEQ4HQ
and the recent ACOT
 suggested scale $\mu = m_c$ for $Q<m_c$, and
$\mu^2=m_c{^2}+0.5Q^2(1-m_c{^2}/Q^2)^n$ for $Q<m_c$ with $n=2$).
A discussion of the
various theoretical schemes for massive charm production is given in
a previous communication~\cite{DXF3}.
Because of a positive correlation between $R$ and  $\Delta xF_3$,
the uncertainty of $\Delta xF_3$ play as a major systematic error
at low $x$ region. However,  for $x>0.1$,  the $\Delta xF_3$ term
is small,
and the extracted values of $R^\nu$ are not sensitive to $\Delta xF_3$.
For a systematic error on the assumed level of
$\Delta xF_3$, we vary strange sea and charm sea 
by $\pm 50$ \% ($\Delta xF_3$ is directly sensitive to the strange sea minus
charm sea). Note that the extracted value of $R$ is larger for a larger
input $\Delta xF_3$ (i.e. a larger strange sea).

\begin{figure}[t]
\centerline{\psfig{figure=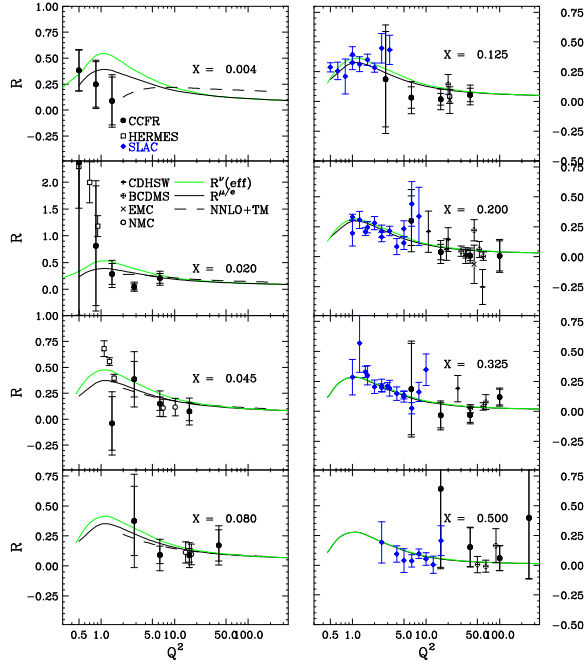,width=3.0in}}
\caption{CCFR measurements of $R^\nu$ as a function of $Q^2$ for fixed $x$,
compared with electron and muon data, with the $R_{world}^{\mu/e}$ and
$R_{eff}^{\nu}$ ($m_c=1.3$) fits, and with the
$R_{NNLOpdfs+TM}^{\mu/e}$)
QCD calculation including NNLO PDFs (dashed line).
The inner errors include both statistical and experimental
systematic errors added in quadrature, and the outer errors include
the additional $\Delta xF_3$ model errors (added linearly).
Also shown are the HERMES results for $R_{N14}^e$ at small $x$ and $Q^2$.}
\label{fig:R}
\end{figure}

The extracted values of $R^\nu$ are
shown in Fig.~\ref{fig:R} for fixed $x$ versus $Q^2$.
The inner errors include both statistical and experimental
systematic errors added in quadrature, and the outer errors include
the additional $\Delta xF_3$ model errors (added linearly).

At the very lowest $Q^2$ values, the model
error is reduced because
all models for $\Delta xF_3$ approach zero around $Q^2 =0.4$.
This is because the strange quark distribution is expected to approach zero
for $Q$ values close to twice the mass of the strange quark. In addition,
the very low $Q^2$ region is below charm
production threshold. Note that the very low $Q^2$ and
low $x$ region is
of interest because it is where HERMES reports~\cite{HERMES} an
anomalous increase in $R^e$ for nuclear targets.

The CCFR  $R^\nu$ values
are in agreement with measurements
of $R^{\mu/e}$,
and  also in agreement with both
the  $R_{world}^{\mu/e}$ and $R_{eff}^{\nu}$ fits.
At low $x$, there are indications that the data may be lower than
these two model predictions.
New QCD calculation~\cite{MRSNNLO}
of $F_L$ and $F_{2}$ including both the NNLO terms and estimates of NNLO PDFs are 
used to determine
$R_{NNLOpdfs+TM}^{\mu/e}$, by including target mass effects. These are
shown as dashed lines in Fig.~\ref{fig:R}.
 There are large uncertainties
in $F_{L}$ from the NNLO gluon distribution at low $x$ for $Q^2<5$ GeV$^2$.
Specifically, for $Q^2=2$ GeV$^2$ and  $0.001<x<0.01$ the NNLO calculation
with NNLO PDFs results in a dip in $F_L$ with $F_L$ approaching zero. 
It is interesting that there is also
a dip in our measured values of $R$ for  $x$=0.019 and $Q^2=3$ GeV$^2$.
However, for $Q^2<2$ GeV$^2$ and  $0.001<x<0.01$ the NNLO calculation
with NNLO PDFs  also yields an unphysical negative value for $F_L$, 
which implies large uncertainties in the calculation. Another recent
QCD  calculation within a $ln(1/x)$ resummation
framework (with resummed 
PDFs) also predicts~\cite{resum} that $R$ at small $x$ and 
low $Q^2$ is lower than $R_{world}^{\mu/e}$.

Also shown are the HERMES electron scattering results in nitrogen.
The HERMES data~\cite{HERMES} for $R$ are extracted
from their ratios for $R_{N14}/R_{1998}$
by multiplying by the values from the $R_{1998}$ fit.
The CCFR data do not clearly show a large anomalous increase at very
low $Q^2$ and low $x$.
It is expected that any nuclear effect
in $R$ would be enhanced in the CCFR iron target
with respect to the nitrogen target in HERMES.  However,
depending on the origin, the effects in electron versus $\nu_\mu$ charged
current scattering could be different.

\end{document}